\def\year{2020}\relax
\documentclass[letterpaper]{article} 
\usepackage{aaai20}  
\usepackage{balance} 
\usepackage{times}  
\usepackage{helvet} 
\usepackage{courier}  
\usepackage[hyphens]{url}  
\usepackage{graphicx} 
\usepackage{graphicx}  
\usepackage{booktabs}
\usepackage{subfigure}
\usepackage{amsmath}
\usepackage[usenames,dvipsnames]{xcolor}
\usepackage{lipsum} 
\usepackage{multirow}
\usepackage{verbatim}

\usepackage{color, colortbl}
\usepackage{fancyhdr}
\usepackage{ctable}
\definecolor{Gray}{gray}{0.9}
\newcolumntype{M}[1]{>{\arraybackslash}m{#1}}
\usepackage{enumitem}
\usepackage{graphicx}
\usepackage{array}
\usepackage{booktabs} 
\usepackage{makecell}
\usepackage{cellspace}
\usepackage{tabularx}

\newcommand{\rev}[1]{\textcolor{black}{#1}}

\newcommand{\crc}[1]{\textcolor{black}{#1}}

\title{The Healthy States of America: \crc{Creating a Health Taxonomy with Social Media}}

\author{Sanja \v{S}\'{c}epanovi\'{c},\textsuperscript{\rm *} Luca Maria Aiello,\textsuperscript{\rm *\rm \dag}
 Ke Zhou,\textsuperscript{\rm *} Sagar Joglekar,\textsuperscript{\rm *} Daniele Quercia\textsuperscript{\rm *\rm \ddag}\\
\textsuperscript{\rm *}Nokia Bell Labs, Cambridge, UK; 
\textsuperscript{\rm \dag}IT University of Copenhagen, DK; 
\textsuperscript{\rm \ddag}CUSP King's College London, UK}

\begin{document}

\maketitle

\begin{abstract}
Since the uptake of social media, researchers have mined online discussions to track the outbreak and evolution of specific diseases or chronic conditions such as influenza or depression. To broaden the set of diseases under study, we developed a Deep Learning tool for Natural Language Processing that extracts  mentions of virtually any medical condition or disease from unstructured social media text. \crc{With that tool at hand, we processed Reddit and Twitter posts, analyzed the clusters of the two resulting co-occurrence networks of conditions, and discovered that they correspond to well-defined categories of medical conditions.} This resulted in the creation of the first comprehensive taxonomy of medical conditions automatically derived from online discussions. \crc{We validated the structure of our taxonomy against the official International Statistical Classification of Diseases and Related Health Problems (ICD-11), finding matches of our clusters with 20 official categories, out of 22.} Based on the mentions of our taxonomy's sub-categories on  Reddit posts geo-referenced in the U.S., we were then able to compute disease-specific health scores. As opposed to counts of disease mentions or counts with no knowledge of our taxonomy's structure, we found that our disease-specific health scores are causally linked with the officially reported prevalence of 18 conditions.
\end{abstract}

\section{Introduction}
To monitor physical and mental health interventions,  National Health agencies such as the Centers for Disease Control and Prevention (CDC) in the U.S. or the National Health Service (NHS) in the U.K. collect and disseminate prevalence data for a broad range of diseases. However, disease prevalence is only part of the story. Such measurements do not paint ``a full picture'' of people's own health experiences. What are the patients' concerns? Which symptoms do they experience? How do these symptoms evolve?

To get a richer picture, some countries conduct periodic health surveys. For example, England regularly runs surveys within the Quality and Outcomes Framework (QOF) to gauge care quality achievements across the entire country~\cite{Gillam}. However, these surveys generally have four main limitations: \emph{1)} they are temporally coarse-grained (they are administered every 3 to 5 years); \emph{2)} they are costly; \emph{3)} they suffer from recall biases given the retrospective nature of recalling past experiences; and \emph{4)} they are administered by doctors and answered by patients who may well reply in a manner that they perceive will be viewed favorably by the doctors themselves~\cite{gkotsis2017characterisation}.

To produce more adequate health assessments, researchers have investigated the linguistic characteristics of content shared on social media. That is mainly because social media platforms have become a source of `in-the-moment' daily exchanges on a variety of subject matters, including health. As such, studying such platforms holds the key to understanding \emph{what concerns patients (rather than doctors)} most~\cite{gkotsis2017characterisation}. The number of patients who use social media to solicit support, to discuss symptoms and remedies, or to simply vent is rapidly increasing, thus resulting in thriving health platforms~\cite{kass2013social,sarasohn2008wisdom}. Such rich crowdsourced information has encouraged researchers to study health-related behaviors at scale~\cite{kass2013social}. Indeed, using social media data, previous research monitored the spreading of infectious diseases~\cite{velasco2014social} and addiction~\cite{Balsamo2019}, and tracked non-communicable conditions such as depression~\cite{de2013predicting,bagroy2017social}, and obesity~\cite{Culotta204twitter}. Recently, risk scores estimated from online data have been proposed even for sexually transmitted diseases~\cite{chan2018online}, and stress~\cite{guntuku2019understanding}.

However, if social media studies are to be blended with official data, three issues need to be addressed. The first two issues were already identified in a 2014 Science article~\cite{Lazer1203}. In that article, Lazer \emph{et al.} analyzed the parable of Google Flu Trend, a Google platform that predicted flu trends based on search queries. The authors considered that particular platform because it was often held up as an exemplary use of big data for health in those days~\cite{mcafee12bigdata,Goel17486}. The authors had identified the two main issues that then led to the platform's demise in 2014, and these issues are still with us today: ``big data hubris'', and ``algorithm dynamics''. \emph{Big data hubris} is \emph{``the often implicit assumption that big data are substitute for, rather than a supplement to, traditional data collection and analysis''}~\cite{Lazer1203}. By contrast, reality has suggested the opposite. It has been found that, by combining Google Flu Trend data with lagged CDC data, over-fitting could have been avoided. Instead, Google's solution purely relied on search queries and, in February 2013,  the platform ended up predicting more than double the proportion of doctor visits related to influenza than CDC's~\cite{butler13}.

The second issue that still needs to be addressed is ``\emph{algorithm dynamics}''. That is to do with whether ``the instrumentation is actually capturing the theoretical construct of interest''~\cite{Lazer1203}. Google's methodology was to find the best matches among 50 million search queries to fit 1152 points derived from the CDC flu data. ``The odds of finding search terms that match the propensity of the flu but are structurally unrelated, and so do not predict the future, were quite high''~\cite{Lazer1203}. That translates into saying that big data was over-fitting the small number of cases. Nowadays,  any machine learning approach applied to large datasets would suffer from the very same problem. Seth Stephens-Davidowitz says that solutions based on big data are often entrapped by what he calls the \emph{curse of dimensionality}: being large, new data sources ``often give us exponentially more variables than traditional data sources and, if you test enough things, just by random chance, one of them will be statistically significant''~\cite{seth18everybody}. If you test enough search queries to see if they correlate with flu incidence, you will find one that correlates just by chance. This problem is found in areas beyond Internet research as well. It has indeed been a general pattern in genetics research. ``First, scientists report that they have found a genetic variant that predicts IQ. Then scientists get new data and discover their original assertion was wrong''~\cite{seth18everybody}.

The third and final issue speaks to the \emph{need for broad health measures}, rather than measures tailored to specific diseases. Many social media studies focus only on very narrow yet important outcomes. Health, however, consists of a much broader range of aspects, certainly including illnesses and diseases, but also encompassing more general health conditions. Despite all symptoms and diseases are connected by a network of complex relationships~\cite{zhou2014human}, most health-related social media studies have so far focused on individual diseases or limited sets of conditions. This is partly due to the historical difficulty in developing text mining models that generalize across multiple health domains. \crc{Because of that, previous work focusing only on the preselected conditions could not automatically discover the complex underlying medical taxonomy expressed by people on social media.}

Our work partly tackles these three issues by: \crc{\emph{1)} automatically discovering the medical taxonomy present in online discussions; \emph{2)} based on the discovered taxonomy,} proposing social media health metrics for a variety of conditions that can be blended with official data; \emph{3)} computing each condition's metric based on the limited set of symptoms related to that condition without over-fitting on, for example, unrelated terms; and \emph{4)} proposing broader health metrics, making it possible to examine multiple conditions simultaneously. In so doing, we made four main contributions:
\begin{itemize}
\item Based on the latest advancements in Deep Learning, we developed a Natural Language Processing tool that can extract mentions of virtually any symptom or disease from unstructured text (\S\ref{sec:extraction}). When applied to standard benchmarks, our approach beats the best performing methods proposed in recent literature and achieves high accuracy on social media data.
\item We applied our extractor of health mentions on 7M+ posts and 130M+ comments authored by geo-referenced Reddit users. The network that emerges from condition co-occurrence within posts and comments represents the first map of general health discussions in social media (\S\ref{sec:network}). Using community detection on this network, we exposed its highly modular structure arranged in 34 top-level clusters and 241 sub-clusters of known medical conditions. When applying the same procedure on 225M tweets, we found a comparable cluster structure but with distinctions that reflect the different nature of the two platforms. \crc{We validated the structure of extracted Reddit taxonomy (the more specialized yet more comprehensive between the two) against the official  International Statistical Classification of Diseases and Related Health Problems (ICD-11), finding that 20 of official categories out of 22 in total are matched by our clusters.} Using the newly found cluster structure from Reddit, we defined several health scores to measure population health from online discussions (\S\ref{sec:HI}).
\item When computed on geo-referenced Reddit posts, disease-specific health scores negatively correlate with 18 corresponding statistics of medical conditions estimated at the level of states in the U.S.\ (\S\ref{sec:HI:validity}). For example, we found \crc{significant} correlations between our mental health score and surveys on mentally unhealthy days ($r=-.45$), our obesity score with diabetes prevalence statistics ($r=-.45$), and our STDs score with the prevalence of syphilis ($r=-.47$). Finally, a more general composite heath score best correlates with self-rated overall health ($r=-0.39$).
\item Moving beyond correlations, we used causal inference and confirmed the causal impact of the prevalence of 12 medical conditions (out of the 14 we considered in total) on the health scores derived from Reddit at the state level in the U.S.\ (\S\ref{sec:causality}).
\end{itemize}

\section{Extracting medical conditions from social media text} \label{sec:extraction}

We developed a Natural Language Processing (NLP) algorithm to extract mentions of medical conditions from text (\S\ref{sec:extraction:NER}), trained it on a dataset that we labeled through crowdsourcing, and applied it on geo-referenced Reddit and Twitter posts at scale (\S\ref{sec:extraction:geoposts}, \S\ref{sec:extraction:application}).

\subsection{NLP medical entity extractor} \label{sec:extraction:NER}

Named Entity Recognition (NER) is the task of extracting entities of interest from text. A NER model identifies $n$-grams that are likely to represent an entity of a given type. In the medical domain, the entities considered are usually symptoms (and associated diseases), and drug names. In this work, we trained an entity extractor to detect medical \emph{conditions} (which include both symptoms and diseases). State-of-the-art NER models are based on Recurrent Neural Networks (RNNs) and contextual embeddings~\cite{jiang2019improved,akbik2018contextual,devlin2019bert}. We implemented a sequence modeling RNN architecture composed by a bidirectional LSTM with a Conditional Random Field layer~\cite{huang2015bidirectional} using RoBERTa contextual embeddings~\cite{liu2019roberta}.

To assess the performance of our model, we trained and tested it using two standard benchmark datasets for medical entity extraction: CADEC~\cite{karimi2015cadec} and Micromed~\cite{jimeno2015identifying}. CADEC contains $1,250$ posts from the AskAPatient forum, all annotated by experts who marked mentions of adverse drug reactions, symptoms, clinical findings, diseases, and drug names (we grouped the first four categories into one category). Micromed contains $734$ tweets annotated in terms of symptoms, diseases, and pharmacological substances. Our method outperformed the state-of-the-art entity extraction approaches in extracting symptoms both on CADEC~\cite{tutubalina2017combination} (F1 score of .78 against a .71 baseline) and Micromed~\cite{yepes2016ner} (F1 score of .74 compared to .59).

The structure of CADEC and Micromed posts is notably different from that of the typical Reddit post. To preserve a high annotation quality when applying our entity extractor to Reddit, we re-trained our model on Reddit data. We created \emph{MedRed}: a new dataset of  Reddit posts labeled with medical entities (symptoms, diseases, and drug names).

\rev{We first sampled $1,980$ posts at random from $18$ subreddits, each dedicated to a specific disease. We then obtained labels for each post through a crowdsourcing task on Amazon Mechanical Turk, which we set up with labeling instructions similar to those used to create the CADEC dataset~\cite{karimi2015cadec}. We restricted the crowdsourcing task only to workers with an approval rate record above 95\%. Given a Reddit post, we asked the workers to copy-paste the symptoms and diseases that they could find in the text. We assigned to the workers batches of four posts, mixed with two additional `control' entries whose medical entities were known to us: one `control' Reddit post with a clearly identifiable symptom, and one entry from CADEC. We discarded the whole batch, if the worker mislabelled the control post, which happened in roughly 21\% of the cases. Each post was shown to 10 workers. In line with previous literature~\cite{lawson2010annotating}, we considered only the list of entities $A_{workers}$ that were independently found by at least two workers. To assess the quality of the crowdsourcing results, for each CADEC entry $i$, we computed the pair-wise agreement between the list of entities extracted by the workers and the ground-truth list of entities $A_{expert}$ extracted by the CADEC experts:}
$$
Agr_{workers, expert}(i) = \frac{match(A_{workers}(i), A_{expert}(i)) }{ \max(|A_{workers}(i)|, |A_{expert}(i)|) },
$$
\rev{where $match$ is a matching function that counts the number of common entries between the two lists. We experimented with two implementations of this function, one that uses `exact string' matching, and one that uses relaxed string matching (e.g., `pain' would be a positive match for `strong pain'). We measured a strict agreement of .62 and a relaxed agreement of .77, which indicates that the quality of annotation from the workers is close to the expert annotations. When training on this data, our entity extractor achieved an F1-score of $.71$ (50\% train, 25\% tuning and 25\% test).}

\subsection{Geo-located Reddit posts} \label{sec:extraction:geoposts}

Reddit is a public discussion website and the fifth most popular website in the U.S. Reddit is structured in roughly 140k independent subreddits (subcommunities) dedicated to a broad range of themes,  including a variety of health and well-being topics (e.g., /Depression, r/HealthyFood, r/Fitness). Users can post new \emph{submissions} to any subreddit, and add \emph{comments} to submissions or to existing comments. From Pushshift, a public collection of Reddit content~\cite{baumgartner2020pushshift}, we gathered all the submissions made during the year 2017, for a total of 96M submissions by 14M users.

To match Reddit discussions with official health data, we focused on users we could locate at the level of states in the U.S. Reddit does not provide any explicit user location, yet it is possible to get reliable location estimates with simple heuristics. Following previous work~\cite{Balsamo2019}, we first selected 2,844 subreddits related to cities or states in the U.S. From those subreddits, we listed the users with at least 5 posts or comments and removed those who posted contributions on subreddits in multiple states. We thus obtained a list of 484,440 users who are likely to be located in one of the 50 U.S.\ states. \rev{In 2017, these users authored 7,162,703 posts, and 134,861,496 comments.}

\begin{figure}
		\centering
		\includegraphics[width=0.49\columnwidth]{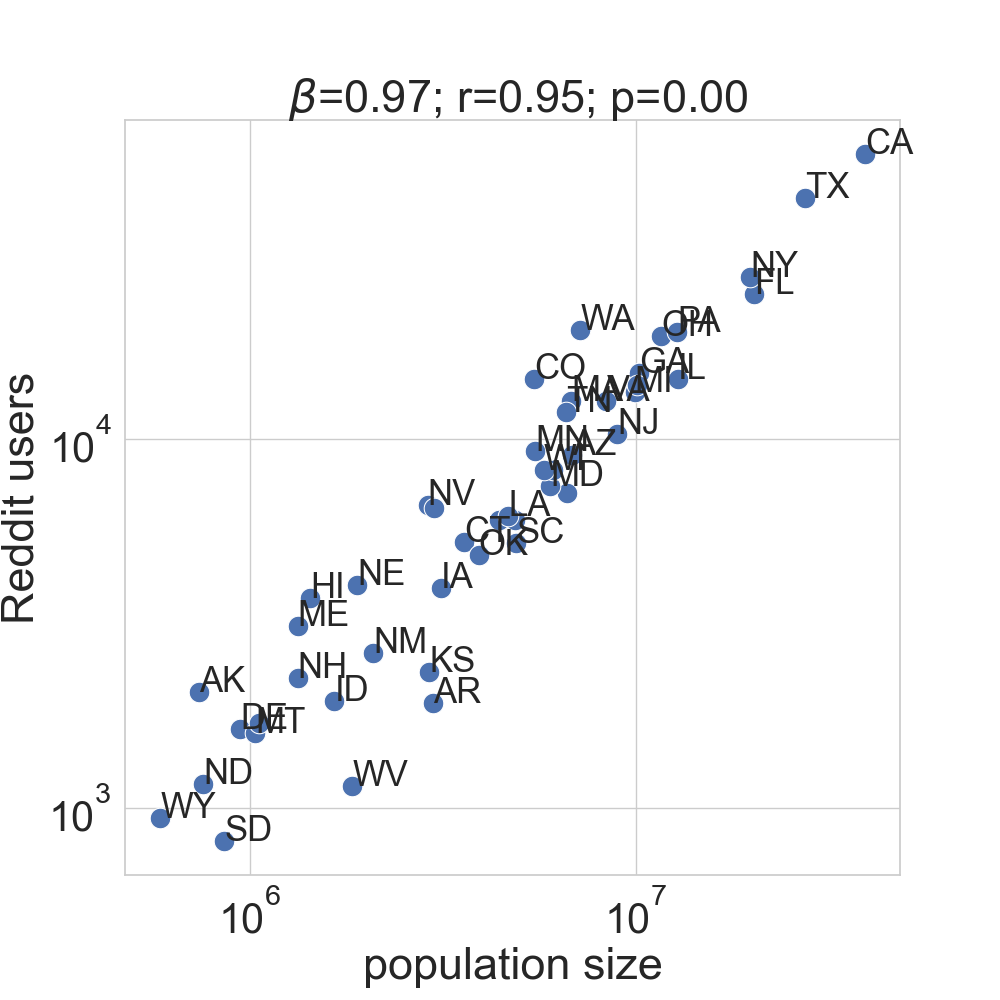}
		\includegraphics[width=0.49\columnwidth]{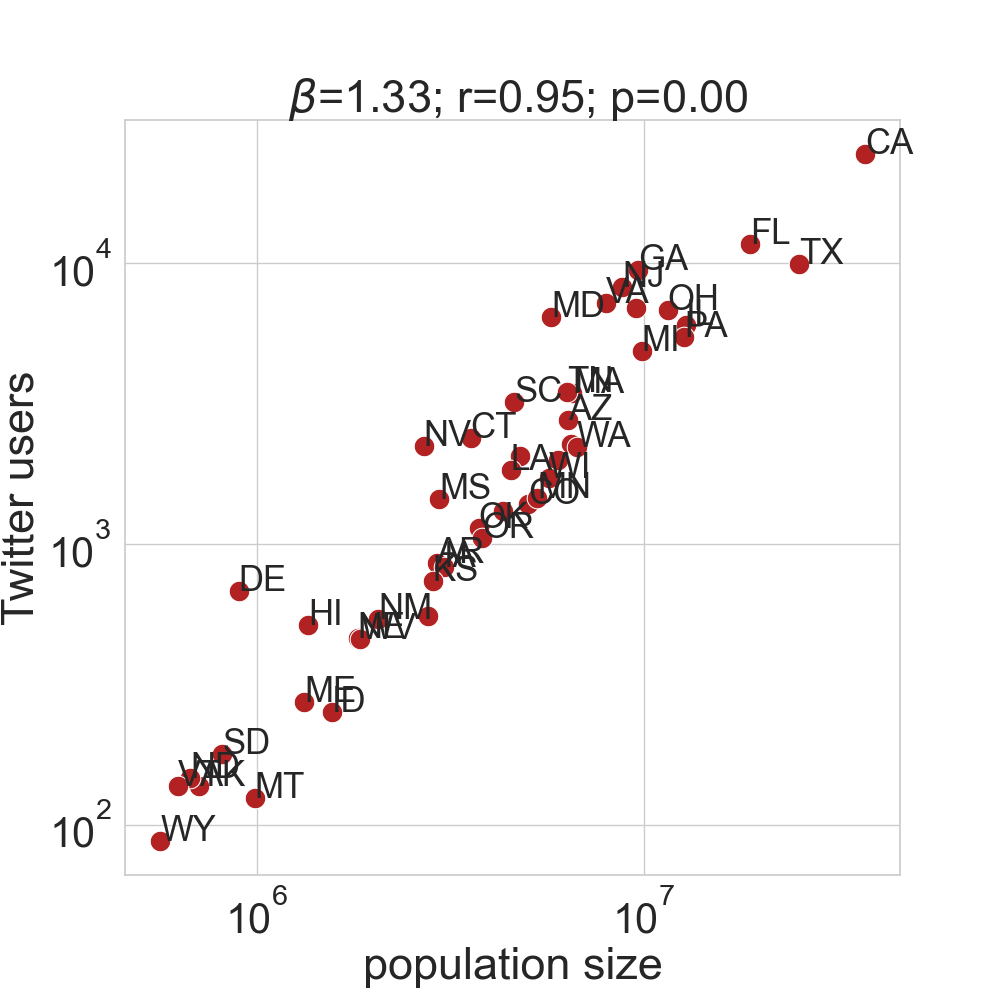}
	\caption{Relationship between the number of Reddit users (\emph{left}), Twitter users (\emph{right}) and state population (\emph{log}-transformed). Pearson correlation for Reddit $r=.95$ and $p<e^{-23}$, and for Twitter $r=.94$ and $p<e^{-21}$.}
	\label{fig:Reddit_user_in_states}
\end{figure}

We checked the representativeness of the data by computing the ratio between the number of Reddit users located in a state and the state's population size as per the 2015 census (the census year closest to the data collection period). We found that this ratio deviated more than two standard deviations from the average for three states: \textit{Mississippi}, \textit{Oregon}, and \textit{Vermont}. After excluding these outliers, the number of Reddit users strongly correlated with population size (Pearson $r=.95$, Figure~\ref{fig:Reddit_user_in_states} left).

\subsection{Geo-located tweets} \label{sec:extraction:geotweets}

\rev{Twitter is a popular micro-blogging service, with more than $300$M active monthly users. On Twitter, users post short messages (tweets) that are shown to their followers. Unlike Reddit posts, tweets can be geo-referenced with the device's GPS coordinates recorded at the time of tweeting. We collected a random sample of 225M tweets posted from the U.S.\ in the year 2010. In our sample, the number of Twitter users across states correlates strongly with census population from the same year (Pearson $r=.94$, Figure~\ref{fig:Reddit_user_in_states} right)}.

\subsection{Medical conditions in social media} \label{sec:extraction:application}
\begin{figure}[t!]
	\centering
		\includegraphics[width=0.48\columnwidth]{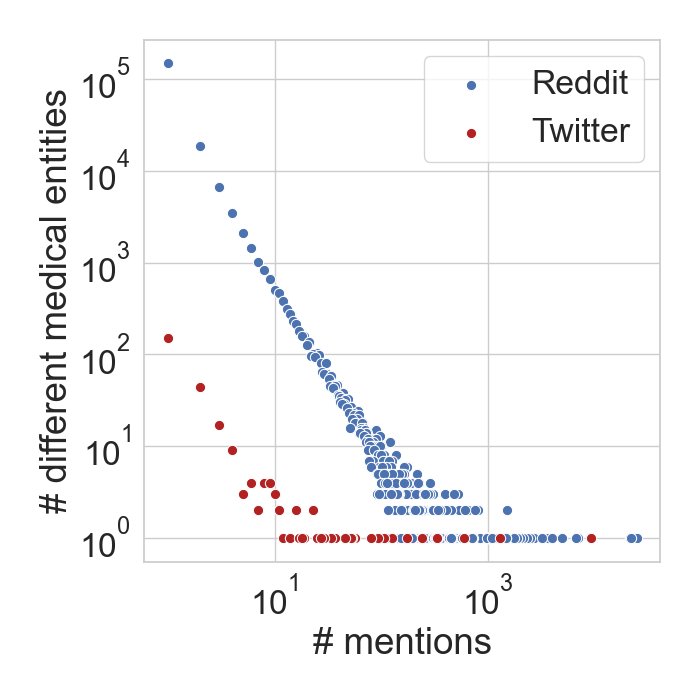}
		\includegraphics[width=0.48\columnwidth]{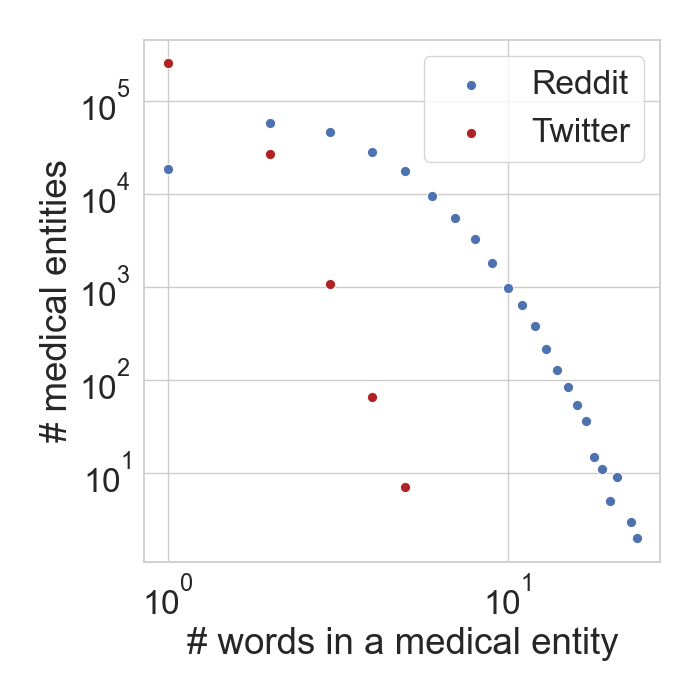}
	\caption{Medical conditions extracted from Reddit \crc{(submissions)} and Twitter: distribution of their mention frequency (\emph{left}), and distribution of number of words per extracted medical condition (\emph{right}).}
	\label{fig:symptom_stats}
\end{figure}

When applied to Reddit, our extractor of medical conditions found 818,656 mentions of medical conditions from 531,081 submissions (7\% of the total), \crc{23,982,372 mentions of conditions from 4,867,759 comments (20\% of the total)}, authored by 180,401 users (more than 37\% of all users). We filtered out submissions \crc{and comments} from a number of subreddits that could introduce undesired topical biases (including subreddits related to animals, fashion, and computers), which left us with 738,152 mentions of 189,456 unique medical conditions in submissions, and \crc{22,787,244 mentions of 869,029 unique medical conditions in comments}. The medical conditions most frequently mentioned were variations of the words \emph{depression, pain, anxiety, cancer}, and \emph{stress.} Their frequency distribution is broad, with the majority of medical conditions mentioned just once (Figure~\ref{fig:symptom_stats}, left). The typical symptom is composed by two words, but we also found more complex conditions with descriptions up to 25 words (Figure~\ref{fig:symptom_stats}, right) such as: \textit{``sharp tight pain in the left side of my chest that also goes into my back and for some reason up into my ear''}.
\rev{In Twitter, our method found 280,177 mentions of medical conditions in 258,245 tweets posted by 108,437 users. These tweets contained 13,261 unique conditions composed by 5 words at most---which is expected, given that tweets are limited to a maximum of 280 characters. The most frequent symptoms include variations of the words \emph{tired, hungry, sick, pain,} and \emph{headache}.}

\section{Classes of symptoms and diseases} \label{sec:network}

To structure the medical conditions extracted from Twitter and Reddit separately, we derived two 
co-occurrence networks (\S\ref{sec:network:infomap}). By hierarchically clustering their nodes, we then obtained two taxonomies (\S\ref{sec:network:taxonomy} and \S\ref{sec:network:taxonomy_twitter}).

\subsection{Clustering the networks of medical conditions
\label{sec:network:infomap}}
We built two co-occurrence networks of the extracted medical conditions, one from Reddit, and the other from Twitter. In these networks, nodes are medical conditions, undirected edges connect pairs of conditions mentioned in the same message (either a submission, comment, or tweet), and edge weights are equal to the number of co-mentions in the same message. These co-occurrence networks capture the semantic relatedness of medical conditions: symptoms or diseases that were often mentioned together are likely to describe the same class of conditions, and in the network, they form a densely-connected cluster of nodes. To find these semantically cohesive clusters, we used network \emph{community detection}. Community detection algorithms partition the network into groups (or \emph{clusters}) of densely connected nodes that are sparsely connected with nodes in other groups. The density of co-occurrence networks is typically high, which negatively impacts the performance of these algorithms. To mitigate this issue, it is standard practice to sparsify the network beforehand. Using noise-corrected backboning~\cite{coscia2017network}---a technique that relies on a statistical null-model to identify and prune non-salient edges---we reduced the Reddit network from $1.8M$ to $1M$ edges, and the Twitter network from $130k$ to $27k$ edges. We focused our analysis on the two giant connected components, which contain $411k$ nodes in the Reddit network, and $6k$ nodes in the Twitter network.

To find clusters in these two networks, we could have used any of the literally thousands of different community detection algorithms that have been developed in the last decades~\cite{fortunato2010community}. Among them, we opted for Infomap~\cite{rosvall2008maps}, a widely adopted algorithm that exhibited very good performance across several benchmarks~\cite{lancichinetti2009community}. Furthermore, Infomap suits our study because: \emph{i)} it extracts a \emph{hierarchical} arrangement of clusters that directly maps to a taxonomy of conditions; and \emph{ii)} it computes \emph{overlapping} clusters, thus identifying conditions that play a role in multiple clusters. 

In the remainder, we will refer to these clusters as \emph{categories} of medical conditions.

\subsection{Taxonomy of medical conditions from Reddit} \label{sec:network:taxonomy}

{\def\arraystretch{1.50}
\begin{table*}[t!]
\tiny
\begin{center}

\begin{tabular}[t]{p{9mm}p{40mm}p{25mm}}

\multicolumn{1}{c}{\textbf{\textit{\textbf{(A)} Level-1}}} & \multicolumn{1}{c}{\textbf{\textit{Level-2}}} & \multicolumn{1}{c}{\textbf{\textit{Example words}}} \\
\Xhline{2\arrayrulewidth}
\rowcolor{Gray} \multicolumn{3}{c}{\textbf{Mental}} \\
mental [06] & isolation, autism, adhd, bipolar, psychosis, severe illness, anhedonia, stress, tic, paranoia, anguish, dyslexia, depression, personality disorder, impulsive behaviour, genetic, psych disorder, fatigue, misophonia, personality, light head, insecurity, dissociation &  wobbly feel, dread, hypomania, autism, suicidal thought\\
anxiety [06] & anxiety & anxiety, anxious, panic attack \\
personality [06] & bpd, dysphoria, narcissistic, antisocial, schizotypal & lack of empathy, sociopathic, manipulative behaviour, abusive behaviour\\

\rowcolor{Gray}\multicolumn{3}{c}{\textbf{Behaviour}} \\
breathing [12] & asthma, fatigue, chest, heart, breathing, active breathing control, inflammations & trouble breathing, severe chest pain, esophagus spasm \\
vomit [21] & vomiting, emetophobia, bugs, pain, gagging & terrible fever, phobic, disgust \\
STDs [01] & stds, yeast, pregnancy, pain &  hiv, syphilis, viral load, losing blood, testicular ache\\
obesity [05] & eating disorders, hunger, weight loss &  obese, overweight, excessive fat, overeating \\
addiction [06] & drugs, porn, alcohol, symptoms & drinking problem, opiates, strong urge, abscess, irritable \\
sleep [07] & hallucinations, pain, traumas, nightmares, apnea, narcolepsis, insomnia, sleepwalking, lack & ptsd, flashback, apnea, snore, wake up every hour \\

\rowcolor{Gray}\multicolumn{3}{c}{\textbf{Body parts}} \\
skin [14] & acne, redness, wrinkles, hyperpigmentation, scalp, aging, dryness, only, spots, bleeding, burns, inflammation, rash, itching, eczema, allergies, bites, herpes, food allergies, soreness, bumps, psoriasis, vitamin, body hair, irritation, scab & pimple, whitehead, flaky, dark spot, ingrown hair, mango allergy \\
ear [10] & tinnitus, dementia, vertigo, vibrations, congestion, noise & ringing in my ear, dizzy, blowing nose constantly \\ 
eye [09] & vision distortion, blurry vision, gallstone, high pressure, eye alignment, blindness, glaucoma, sweating, light sensitivity, strain, hypertension, aneurysm, migraine & eye pressure, spatially aware, nearsighted\\
heart [11] & palpitations, irregular, tachycardia & irregular heartbeat, poor concentration \\
spine [08] & multiple sclerosis, neurogenerative, hernia &  tingling, lesion, difficult to lay \\
back [15] & pain, sciatica, arthritis, lower, stiffness, dullness & hip pain, muscle stiffness, unable to sit up straight\\
reproductive [16, 17] & stones, infections, clots, lupus, bladder & shave, pain with sex, extremely bloated\\

\end{tabular}
\quad
\begin{tabular}[t]{M{9mm}M{40mm}M{25mm}}

\multicolumn{1}{c}{\textbf{\textit{Level-1}}} & \multicolumn{1}{c}{\textbf{\textit{Level-2}}} & \multicolumn{1}{c}{\textbf{\textit{Example words}}} \\
\Xhline{2\arrayrulewidth}
\rowcolor{Gray}\multicolumn{3}{c}{\textbf{Conditions}} \\
cancer [02] & cancer, gout, skin, lymphoma, lumps, genitals, digestive, lymphnodes, bones & discolored skin, swollen lymphnode, back ache, terminally ill, gnarly bruise \\
infective [01] & sepsi, heart, fever, overdose, penumonia, mosquito-borne, measles, blood, pain, confusion & highly viral, dark mucus, sweating and cough, with knuckle, blood clot\\
influenza [01] & viral, flu, yellow fever & increased temperature, loss of appetite \\
diabetes [05] & diabetes, cataract, metabolic syndrome, vision, brain & nebula, brain fog, low blood sugar, lost pigment\\
parkinson [08] & parkinson & tremor, jittering \\
injuries [22] & body, broken, nagging, traumas, head, disorientation & concussion, skull fracture, opiates, sleeping difficulties  \\
parasites [01] & lyme, fungi, fatigue, sleepiness & debilitating fatigue, fungal infection, mold, dark spot \\
epilepsy [08] & seizure, spine, paralysis & spaced out feel, numb, muscle twitch\\

\rowcolor{Gray}\multicolumn{3}{c}{\textbf{Demographics}} \\
female [16] & pcos, hair loss, vagina, cyst, endometriosis, pelvic, ovaries, spasm, weight/swelling, menopause & hot flash, irregular period, swollen, painful cyst, vaginismus\\
infants [18] & reflux, ppd, breast, teeth & spitting up, clogged duct, nipple damage, screaming, mentally drained \\
elderly [-] & arthritis, prostate, hernia & urinary issue, cystitis, struggling to walk\\
pregnancy [18, 19] & birth complications, contractions, pms, shake/ache & regular contractions, bleeding, painful cramp, dilated cervix \\
developmental [20] & birth defects, down syndrome, genetic, edema, preeclampsia, cystic fibrosis & absent nasal bone, unable to digest food, respiratory distress \\

\rowcolor{Gray}\multicolumn{3}{c}{\textbf{Systems}} \\
nervous [08] & migrain, stroke, nerve pain, hemicrania, neck pain, persisting hallucinations & vessel occlusion, allodynia, cephalgia, photosensitive \\
respiratory [12] & cough, ear infection, sinus, sneezing, head, bronchitis, dryness, throat, nose, abdomen & sniffle, lingering cough, runny nose, tight airways, sore throat, abdominal discomfort, yellowish with cough \\

autonomic [-] & hypermobility, fibromyalgia, dysautonomia, erythema, patellofemoral, vasovagal, spasms, severe disfunctions & hard skin, spasm, fainting, arrhythmia \\
digestive [13] & bloating, chron, hemorroid, irritation, bowel inflammation, celiac, constipation, stomach, diarrhea, gastritis, flu & flare, trouble pooping, anal fissure, allergic to gluten \\
tyroid [05] & hypothyroidsm, burning mouth, hashimoto, infections, gastroparesis & lose my hair, growling stomach, swollen tyroid \\

\end{tabular}

\begin{tabular}[t]{M{170mm}}
\Xhline{2\arrayrulewidth}
\rowcolor{Gray}\textbf{(B) \crc{Medical conditions belonging to multiple communities}} \\
cold, itching, inflammation, oily skin, insecure, heart disease, motion sick, blackhead, trouble sleeping, leukemia, runny nose, paralysis, flashback, hearing loss, agitated, confused, extreme anxiety, opioid addiction, tunnel vision, strep throat, munching, health anxieti, dry eye, insane, chronic fatigue, lesion, pale, mental block, warp, losing weight, ovarian cyst, period cramp, celiac diseas, queasy, irregular period, high anxiety, injury, low blood sugar, no energy, postpartum depression, blurry vis, sniffle, sleepless, dry patch, trouble falling asleep, neurot, abnorm, incontinent, dehydrated skin, mentally challenged, mild cramp, emotional stress, hypersensitivity, heat, cloudy, poor sleep, low self confidence, light spot, dark skin, visual snow, shyness, urge, knees hurt, dri, leakage, itchy scalp, uneven skin tone, blood stain, lack motivation, emotional trauma\\
\end{tabular}

\begin{tabular}[t]{M{170mm}}
\rowcolor{Gray}\textbf{(C) ICD-11 categories} \\

\textbf{[01]} Certain infectious or parasitic diseases; \textbf{[02]} Neoplasms; \textbf{[03]} Diseases of the blood or blood-forming organs; \textbf{[04]} Diseases of the immune system; \textbf{[05]} Endocrine, nutritional or metabolic diseases; \textbf{[06]} Mental, behavioural or neurodevelopmental disorders; \textbf{[07]} Sleep-wake disorders; \textbf{[08]} Diseases of the nervous system; \textbf{[09]} Diseases of the visual system; \textbf{[10]} Diseases of the ear or mastoid process; \textbf{[11]} Diseases of the circulatory system; \textbf{[12]} Diseases of the respiratory system; \textbf{[13]} Diseases of the digestive system; \textbf{[14]} Diseases of the skin; \textbf{[15]} Diseases of the musculoskeletal system or connective tissue; \textbf{[16]} Diseases of the genitourinary system; \textbf{[17]} Conditions related to sexual health; \textbf{[18]} Pregnancy, childbirth or the puerperium; \textbf{[19]} Certain conditions originating in the perinatal period; \textbf{[20]} Developmental anomalies; \textbf{[21]} Symptoms, signs or clinical findings, not elsewhere classified; \textbf{[22]} Injury, poisoning or certain other consequences of external causes\\

\end{tabular}

\end{center}
\caption{\rev{(\emph{A}) The taxonomy of medical conditions extracted from Reddit, arranged in two levels, with some examples of individual conditions. The names of the level-1 and level-2 categories were assigned by the authors after manual inspection. We manually arranged the top-level categories into six coherent themes. (\emph{B}) A selection of the most frequent conditions that belong to multiple categories. (\emph{C}) The list of  top-level categories from International Classification of Diseases (ICD-11) by the World Health Organisation (WHO).}}
\label{tab:taxonomy}
\end{table*}
}

Table~\ref{tab:taxonomy} summarizes the 34 level-1 categories of medical conditions, and the 241 level-2 categories found by Infomap on the Reddit co-occurrence network. These categories cover a wide range of medical conditions. We manually named the level-2 categories after inspecting the 50 most frequently used words they contained; we then manually named the level-1 categories based on the level-2 categories they contained. Finally, we manually grouped the level-1 categories into six main themes (grayed rows in Table~\ref{tab:taxonomy}). That is, symptoms associated with: mental health; individual body parts (e.g., eyes); systems of the human body (e.g., digestive system); specific demographics (e.g., women, elderly);  various behaviors (e.g., eating); or specific conditions (e.g., diabetes, cancer).

To assess the breadth of our taxonomy and to test whether its categories cover well-studied medical conditions, we compared it to the official International Classification of Diseases\footnote{\url{https://icd.who.int}} (ICD-11) of the World Health Organization (WHO), which contains 22 top-level disease categories, further split into sub-categories at multiple hierarchical levels. In ICD, diseases are organized mainly based on the body parts they concern. We matched our level-1 categories to the top-level ICD categories by simply searching the level-1 category on ICD. Out of our 34 level-1 categories, \crc{as many as 32 found a match} (Table~\ref{tab:taxonomy}). Those that did not span multiple ICD categories; for example, our \emph{elderly} category contains conditions frequent among elderly people; yet, since these conditions affect different parts of the body, they are listed across multiple ICD categories. Still, out of the 22 ICD categories, \crc{20 are represented in our taxonomy}, and that makes it the most extensive data-driven categorization of medical conditions.

\begin{figure}
		\centering
		\includegraphics[width=0.50\textwidth]{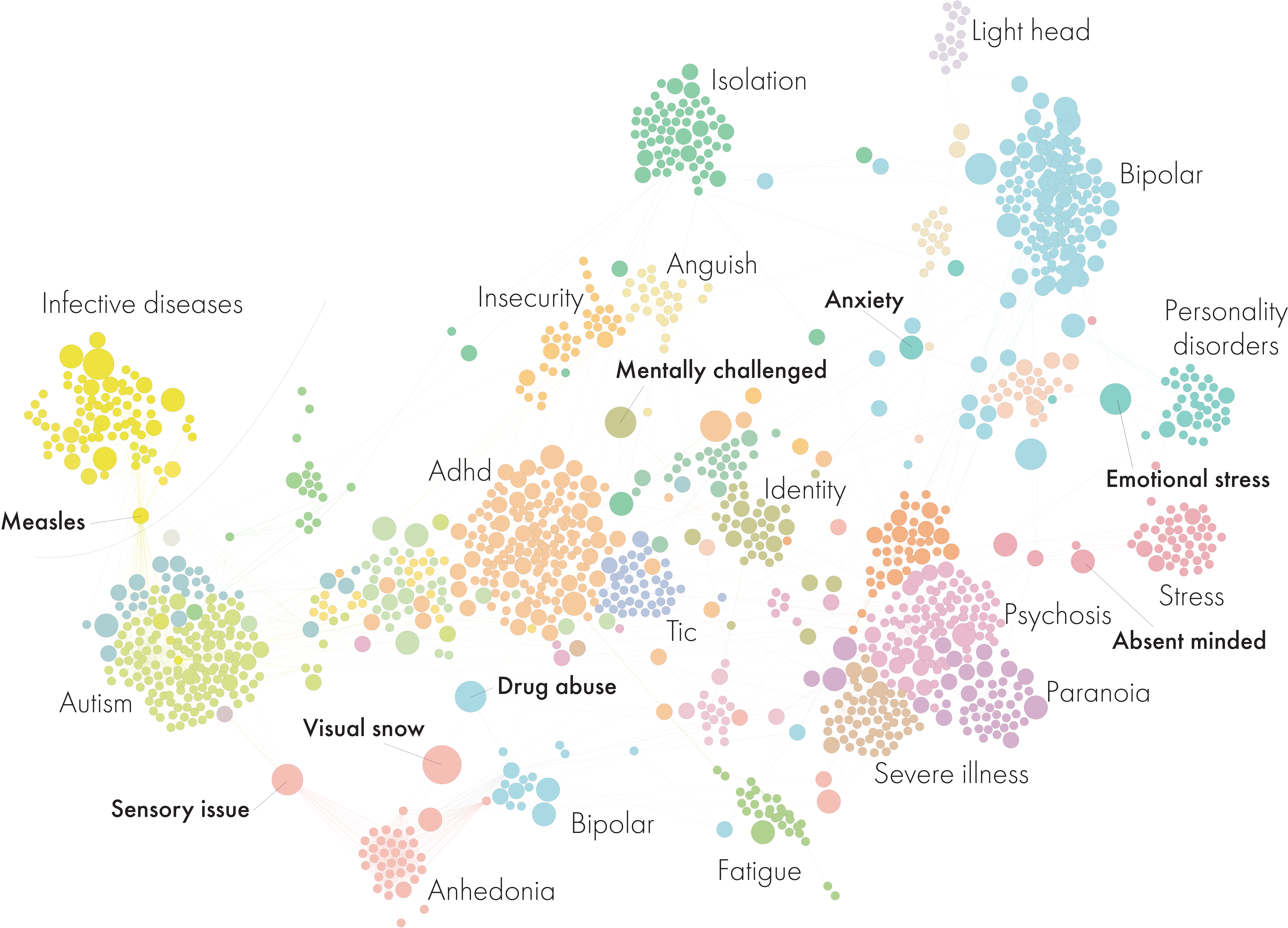}
	\caption{Co-occurrence graph of symptoms in the mental health category. Colors represent level-2 categories and node size is proportional to the number of level-1 categories the nodes belong to. The names of some categories are reported. The names of some nodes that belong to multiple categories is reported in bold. The infective diseases category belongs to a different level-1 category.}
	\label{fig:mental_graph}
\end{figure}

Beyond individual categories, we analyzed individual conditions that occur in multiple categories---at the bottom of Table~\ref{tab:taxonomy}, we listed a selection of the most frequent ones. Most of these conditions are generic symptoms (e.g., itching), \crc{and might appear frequently together with other conditions or symptoms}~\cite{neale1995models}. \crc{In our network, these symptoms are usually not tightly embedded within any of the categories but they tie together different categories}. To illustrate that, Figure~\ref{fig:mental_graph} shows the network of medical conditions within the category of mental health. This network is organized in several, well-separated level-2 categories (e.g., autism, anhedonia), which are often bridged by \crc{symptoms that belong to multiple top-level categories (e.g., sensory issues)}. For example, the \emph{autism} sub-cluster is linked to the \emph{anhedonia} sub-cluster through the medical condition of \emph{sensory issues}---indeed, sensory issues often accompany both autism and anhedonia (the inability to feel pleasure)~\cite{bogdashina2016sensory}. 

Sometimes, cross-category associations emerge from common misconceptions discussed in Reddit. As an example, we show in Figure~\ref{fig:mental_graph} (left-hand side) that \emph{autism} is linked to the top-level community of \emph{infective diseases} through the node \emph{measles} (an infective disease commonly prevented with a vaccine), as a result of the widespread hoax that autism is caused by vaccines~\cite{taylor1999autism}.

\subsection{Taxonomy of medical conditions from Twitter} 
\label{sec:network:taxonomy_twitter}

{\def\arraystretch{1.50}
\begin{table*}[t!]
\tiny
\begin{center}

\begin{tabular}[t]{p{9mm}p{40mm}p{25mm}}

\multicolumn{1}{c}{\textbf{(A)} \textbf{\textit{Level-1}}} & \multicolumn{1}{c}{\textbf{\textit{Level-2}}} & \multicolumn{1}{c}{\textbf{\textit{Example words}}} \\
\Xhline{2\arrayrulewidth}
\rowcolor{Gray} \multicolumn{3}{c}{\textbf{Mental}} \\
mental [06] & personality disorder, depression, ASD, alzheimer, mental illness, panic, anxiety, restless, pregnancy depression &  autism, adhd, disabled, mental, aspergers, ocd, fatigue, dementia, anxiety attack\\
eating disorder [06] & eating disorder & season, nerve pain,  boredom, hungover, appetite\\
\\

\rowcolor{Gray}\multicolumn{3}{c}{\textbf{Behavior}} \\
obesity [05] & obese, complications &  weight gain, heart health, weight loss, thyroid, weary\\
sleep [07] & insomnia & insomnia, mental decline, insomniac, apnea symptom, acute condition \\

\rowcolor{Gray}\multicolumn{3}{c}{\textbf{Systems}} \\
nervous [08] & nervous & syringomyelia, chiari malformation, ventricle syndrom, mental disturb, aneurysm\\
respiratory [12] & asthma, bronchitis, allergies & cough, flu, sinus infect, short, ear infection, breath\\
autoimmune & lupus, scleroderma & lupus, fluid, chronic disease, pain relief, sluggish, nightmare, scleroderma, insomnia, coma \\
digestive [13] & metabolic syndrome, diabetes mellitus & diabetes, hypocalcemia \\
circulatory [11] &  cardiovascular, heart & heart failure, hemorrhagic telangiectasia, hemolytic disease\\
genitourinary [16] & bladder, infectious & bladder cancer, lymphoma, menopause \\ 
\end{tabular}
\quad
\begin{tabular}[t]{M{9mm}M{40mm}M{25mm}}
\multicolumn{1}{c}{\textbf{\textit{Level-1}}} & \multicolumn{1}{c}{\textbf{\textit{Level-2}}} & \multicolumn{1}{c}{\textbf{\textit{Example words}}} \\
\Xhline{2\arrayrulewidth}
\rowcolor{Gray}\multicolumn{3}{c}{\textbf{Conditions}} \\

unwell [21] & tiredness, headache, sore, allergies & {tire, hungri, sick, headach, stress} \\
cancer [02] & breast cancer, skin cancer, colon cancer & breast cancer, skin, colon\\
infective [01] &  fever, mump, hiv, ecoli, h1n1, swine flu, ear, malaria, cholera, salmonella, foodborne illness & fever, flu, sinus headache, tropical depression, humid\\
diabetes [05] & diabetic eye disease  & t-cell, cardiac arrhythmia \\
arthritis [15] & arthritis & osteoarthritis, inflammation, rheumatoid arthritis \\
migraine [08] & headache, migraine & migraine symptom, migrainey, stressed \\

allergies [04] & tiredness, cough & diabetic allergy, chest painsheart attack, sick bloody nose, hangoverish, allergiesasthma \\
leukemia [02] & chronic lymphocytic leukemia  & lymphoma, leukemia, tremor, tension headache\\
sickle cell anemia [03] & sickle cell anemia & sicklecell, pain, sickle cell, excruciating pain sickle cell disease \\
ibs [13] &  irritable bowel syndrome & diarrhea, diabetes remedy, psychiatric condition,chronic malnutrition\\
\rowcolor{Gray}\multicolumn{3}{c}{\textbf{Body parts}} \\
ear [10] & earache & ear ache, bacterial meningitis, tireddd, sniffle, ear hurt \\

\end{tabular}

\begin{tabular}[t]{M{170mm}}
\Xhline{2\arrayrulewidth}
\rowcolor{Gray}\textbf{(B) Medical conditions belonging to multiple communities} \\
depression, tb, whooping cough, hunger, bloat, migraine, anxiety attack, heat rash allergy, aortic tear, salmonella, tremor, kidney stone, discomfort, back pain, concussion, fever, hive, seizure, breath, sinus infection, insomnia, flu \\
\end{tabular}

\end{center}
\caption{\rev{(\emph{A}) The taxonomy of medical conditions extracted from Twitter, arranged in two levels, with some examples of individual conditions. The names of the level-1 and level-2 categories were manually assigned by the authors after content inspection. We also manually arranged the top-level clusters into six coherent themes. (\emph{B}) A selection of the most frequent \emph{co-morbid} conditions that belong to multiple categories.}}
\label{tab:taxonomy_twitter}
\end{table*}
}

The taxonomy extracted from the Twitter network contains 21 level-1 categories, and 53 level-2 categories that match 14 out of the 22 main ICD categories (Table~\ref{tab:taxonomy_twitter}). The Twitter taxonomy exhibits some similarities with Reddit's (10 out of the 21 top-level categories match those from the Reddit taxonomy), but the two differ both in scope and focus. The Twitter taxonomy covers fewer diseases, and contains categories about general unwellness and very common conditions (e.g., migraine, allergies). These dissimilarities reflect the difference between the two platforms: Reddit is a specialized knowledge-exchange platform containing several forums dedicated to specific conditions~\cite{medvedev2017anatomy,de2014mental}, whereas Twitter is a general-purpose microblogging platform whose strict format limitation of 280 characters per message restricts the possibility of in-depth conversations. Our method for extracting medical conditions adapts well to both Reddit and Twitter: it was able to identify meaningful classes of symptoms and diseases in both, despite the stark differences between them. In the remainder, we focus on Reddit, as it has the most comprehensive taxonomy.

\section{Health Scores} \label{sec:HI}
We leveraged our categories of medical conditions to define health scores that we later used to estimate the prevalence of different diseases across states in the U.S. Given the set of conditions $S_i \in S$ in category $i$, and a user $u$ resident in location (state) $l$, we considered the set of conditions $S_i(u) \in S_i$ that user $u$ has mentioned. We determined mentions by directly applying our extractor of medical conditions (\S\ref{sec:extraction:NER}) to both Reddit posts and comments. We then computed the weighted fraction of users in location $l$ who mentioned any condition of category $i$ :
\begin{equation}
f_{i}^{\rho}(l) = \frac{ 1 }{|U_l|} \left( \sum_{u \in U_l} (\max(\{c_{pr}(s), \forall s \in S_i(u) \})^\rho \right)
\label{eq:fu2}
\end{equation}
where $U_l$ is the set of all users in state $l$,  $S_i(u)$ is the set of conditions in category $i$ that user $u$ mentioned, $c_{pr}(s)$ is the Page Rank centrality of condition $s$~\cite{page1999pagerank}, and $\rho$ is equal to either zero or one depending on whether Page Rank centrality is used or not. When $\rho = 0$, the centrality value is discarded, and $f_{i}^{\rho=0}(l)$ becomes simply the fraction of users in $l$ who mentioned conditions in category $i$. By weighting the medical conditions by their centrality on the co-occurrence network, we wanted to give more importance to those that best represent the category they are in. We experimented with variations of Equation~(\ref{eq:fu2}), using sum or average as aggregation functions instead of maximum, and using harmonic centrality or degree centrality instead of PageRank~\cite{boldi2014axioms}. Equation~(\ref{eq:fu2}) is the setup that yielded the health scores that best correlated with official health statistics (\S\ref{sec:HI:validity}). Given these (weighted) fractions, we computed a health score $H_{i}^l$ for category $i$ at location $l$:
\begin{equation}
H^{l}_{i} = - \frac{(f^{\rho}_{i}(l) - \mu_{i})}{\sigma_{i}} 
\label{eq:HI}
\end{equation}
where $\mu_{i}$ and $\sigma_{i}$ are the mean and standard deviation of $f^{\rho}_{i}$ across all locations. The minus makes it possible to have positive values of $H^{l}_{i}$ representing ``healthy'' areas where the fraction of people mentioning conditions in $i$ was lower than average, and negative values representing areas where those symptoms were mentioned more frequently than expected. Similar formulas were used in the past to measure regional happiness and mental well-being from social media posts~\cite{kramer2010unobtrusive,bagroy2017social}.

When calculating a score across locations, we applied a standard outlier removal procedure that excluded the locations in which the ratio $f^{\rho}_{i}(l)$ deviated more than two standard deviations from the overall mean~\cite{kramer2010unobtrusive}. In practice, in all our experiments, this resulted in removing one state (South Dakota). As we previously removed three states because of the lack of representativeness of their user base (\S\ref{sec:extraction:geoposts}), this left us with 46 states in total.

For each of these states, we computed three health scores: $H^l$, $H^l_{i}$, and $H^l_{c}$. We defined them based on three sets of medical conditions: the full set of conditions from all categories; the conditions in category $i$; and the set of most-central conditions on the co-occurrence network (top $5\%$ in the PageRank distribution), respectively.

We compared our health scores with a measure based on simple word matching, inspired by dictionary-based approaches such as LIWC (Linguistic Inquiry and Word Count)~\cite{pennebaker2001linguistic}. We created Dis-LIWC (Disease Linguistic Inquiry and Word Count): a dictionary of symptoms that are known to be frequently associated with certain diseases. We compiled this list by collating four existing sources: the Human Disease Network (HDN)~\cite{goh2007human}, a dataset of over 100K symptom-disease pairs frequently co-mentioned in publications indexed by PubMed; and the set of symptoms that appeared in the `disease description' panels on MedScape\footnote{\url{https://www.medscape.com}}, WebMed\footnote{\url{https://www.webmd.com}}, and Wikipedia. We were able to build a comprehensive Dis-LIWC for 10 conditions, including \emph{depression}, \emph{diabetes}, \emph{asthma}, and \emph{rheumatoid arthritis}. After applying Dis-LIWC to our set of geo-referenced Reddit posts, we computed $H^{l}_{i}$ for these 10 disease categories using Formula~(\ref{eq:HI}), where $\rho=0$ (equivalent to simply counting the occurrences of Dis-LIWC words).

\section{Validity of our Health Scores} \label{sec:HI:validity}
To verify that our health scores are not just proxies for activity levels, we correlated them with four variables obtained for each state in the U.S.: population estimates for the year 2017 produced by the United States Census Bureau\footnote{https://www.census.gov}; the number of Reddit users per 1000 residents\footnote{https://www.reddit.com/user/epickillerpigz}; Reddit adoption calculated as the number of Reddit users in our dataset divided by the census population; and the total number of Reddit users who mentioned medical conditions. None of our health scores correlated with any of those variables (Pearson $r \in [0,0.03]$, $p > 0.1$). After this preliminary check, we proceeded to validate our health scores against state-level health outcomes.

\subsection{External Validity Against Official Statistics}
We collected health statistics from the Center for Disease Control and Prevention (CDC)---a leading public health institute---and from the Substance Abuse and Mental Health Services Administration (SAMHSA), both of which regularly publish health statistics in the U.S. To best match our level-1 categories, from CDC, we gathered state-level prevalence statistics for arthritis, asthma, and self-reported `mentally unhealthy days' and `poor health', compiled between 2016 and 2017. From SAMHSA, we collected statistics on the prevalence of: mental illnesses, abuse of different substances (e.g., heroin), conditions linked to metabolic syndrome (e.g., diabetes prevalence), and Sexually Transmitted Diseases (STDs). All SAMHSA statistics were compiled between 2017 and 2018. In total, we collected 18 health statistics (``Official statistic'' column in Table~\ref{tab:correlations}).

With these statistics at hand, we could test two hypotheses:

\vspace{3pt}\noindent \rev{\textbf{H1:} The prevalence of a specific health condition $i$ measured by official statistics negatively correlates with the corresponding health score $H_{i}^l$;}
	
\vspace{3pt}\noindent \rev{\textbf{H2:} Poor self-reported general health negatively correlates with our general health scores $H^l$ and $H_{c}^l$.}

To find the conditions upon which to calculate the health scores $H_{i}^l$ in \textbf{H1}, we manually parsed all our level-1 categories in our taxonomy to find the best match. In the majority of cases, the mapping was straightforward (e.g., HIV prevalence with the STDs category), except for statistics on arthritis, cocaine use, and heroin use, which do not have a direct mapping. We mapped arthritis to our category \emph{elderly} (which contains the level-2 category \emph{arthritis}), and cocaine and heroin use to our category \emph{infections} (which contains a level-2 category \emph{overdose}).

The correlation results summarized in Table~\ref{tab:correlations} suggest two key insights. First, the Dis-LIWC baseline performs poorly, yielding no statistically significant correlation: simple word-matching strategies do not capture the relationship between online discussions and health outcomes. \crc{Second, considering information from the co-occurrence network in the form of the centrality scores of individual conditions ($\rho=1$), strengthens the correlations compared to using mention counts only ($\rho=0$). Indeed, the centrality-weighted scores achieve stronger correlations with both the statistics on specific conditions ($-0.29 \leq r \leq -0.47$, which supports \textbf{H1}), and the statistic on overall poor health ($-0.33 \leq r \leq -0.39$, which supports \textbf{H2}).}

\begin{table}[ht!]
	\scriptsize
	\begin{center}
		\setlength\tabcolsep{1mm}
		\begin{tabular}{p{14mm}p{26mm}p{7mm}p{8mm}p{7mm}p{7mm}p{5mm}} 
			\hline
			\textbf{Health score}  & \textbf{Official Statistic} & \textbf{$r_{\rho=0}$}  & \textbf{$r_{\rho=1}$} &  \textbf{$r_{liwc}$}  & \textbf{ATE} & \#Conf. \\
			\rowcolor{Gray} $H_{S_i}^l$ & \multicolumn{6}{l}{\textbf{Mental Health}}  \\
			mental  & Mentally Unhealthy Days & -.31* & -.45** &-.06 & -.10*  & 9 \\
			mental  & Mental Illness &  -.23* & -.30*	&- .01 & -.01  & 6 \\
			\rowcolor{Gray} $H_{S_i}^l$ & \multicolumn{6}{l}{\textbf{Substance Abuse}} \\
			breathing  & Cigarette Use & -.31* & -.29*	& --- & -.10* & 7 \\
			infections & Cocaine Use & -.25 & -.29*  & --- & -.06* & 5 \\
			infections & Heroin Use & -.30* & -.43**&  --- & -.09* & 5 \\
			\rowcolor{Gray} $H_{S_i}^l$ &  \multicolumn{6}{l}{\textbf{Metabolic syndrome}} \\
			obesity & High Cholesterol Prev. & -.29*   & -.46*** &-.12  & -.04* & 8 \\
			obesity & High Blood Pressure  & -.26 & -.45*** & --- & -.07* & 4 \\
			obesity & Mortality Cardiovascular & -.19 & -.39** & -.01  &  -.01 & 4 \\
			obesity & Mortality CHD  & -.16 & -.47*** & -.03 & -.07* & 5 \\
			obesity & Mortality Heart Disease  & -.21 & -.39** & -.02  & -.01 & 5 \\
			obesity & Overweight  & -.01 & -.33* & --- & -.07* & 4\\
			obesity & Diabetes Prev.  & -.25 & -.45*** & .02 & -.02 & 5 \\
			\rowcolor{Gray} $H_{S_i}^l$ &  \multicolumn{6}{l}{\textbf{Specific Diseases}} \\
			elderly & Arthritis  & -.45** & -.47*** & -.06 & -.05* &  5 \\ 
			breathing & Asthma & -.33* & -.42** & -.13 & -.06* & 3 \\
			\rowcolor{Gray} $H_{S_i}^l$ &  \multicolumn{6}{l}{\textbf{STDs}} \\
			STDs & HIV prevalence  & -.23 & -.43** & .14 & -.06* & 6 \\
			STDs & AIDS prevalence & -.22  & -.41** & .11 & -.05* &  5 \\
			STDs & Prim. and Sec. Syphilis & -.22  & -.47*** & .23 & -.03* & 7 \\
			STDs & Early Latent Syphilis  &   -.28* & -.39** & .10 &-.12* & 5 \\
			\rowcolor{Gray} $H_{S}^l$ & \multicolumn{6}{l}{\textbf{All Conditions}} \\
			all & Poor Self-rated Health & -.34** & -.33*  & --- & -.19* & 8 \\ 
			\rowcolor{Gray} $H_{S_c}^l$ & \multicolumn{6}{l}{\textbf{Most Central Conditions}} \\
			most central & Poor Self-rated Health & -.38** & -.39** & --- & -.12* & 7 \\ 
			\hline
		\end{tabular}
	\end{center}
	\caption{The link between health scores computed at the level of U.S.\ states and official health statistics. Pearson correlations $r$ are reported for $\rho = 0$ (medical conditions with equal weighting), $\rho = 1$ (medical conditions weighted by their centrality on the co-occurrence graph), and for the Dis-LIWC baseline. P-values classes are also reported (* $p<0.1$, ** $p<0.05$, *** $p<0.01$). The average treatment effect (ATE) of the health statistics on the best performing health scores ($\rho = 1$) are also reported and marked with * when the ATE's confidence interval lies entirely below zero. \#Conf. represents the number of confounders selected for the causal analysis.}
	\label{tab:correlations}
\end{table}

\section{Causal Link Between Health Outcomes and Health Scores} \label{sec:causality}

To go beyond correlation analysis, we set up a causal inference framework (\S\ref{sec:causality:method}) to estimate the causal effect of the prevalence of different diseases on their mentions on Reddit captured by our health scores (\S\ref{sec:causality:results}).

\subsection{Estimating Causality through Matching} \label{sec:causality:method}

In experimental studies, \emph{Randomized Control Trials} (RCTs) are used to estimate the causal effect of a \emph{treatment} on an \emph{outcome}. RCTs select random subjects, assign a treatment to a subset of them, and finally measure the differences in the outcome between the treated and untreated groups. In observational studies, RCTs are not applicable; instead, \emph{matching} techniques are often used to infer causation. Matching works by pairing subjects that were exposed to different either the treatment or outcome but were comparable in terms of \emph{confounding variables}---those factors that may affect being assigned to the treatment or to the control group, or that may affect the outcome---are comparable. The magnitude of the causal effect is then estimated with the \emph{Average Treatment Effect} (ATE), namely the average difference of the outcome variable between paired subjects:
\begin{equation}
ATE = \frac{\sum_{(s_0,s_1) \in M} y(s_1) - y(s_0)}{|M|},
\label{eqn:ate}
\end{equation}
where $y$ is the outcome, $M$ is the set of paired subjects, and $s_1$ and $s_0$ are two comparable subjects, one ($s1$) in the treatment group, and the other ($s_0$) in the control group.

In our setup, subjects are the 46 U.S.\ states we considered, the treatment is a binary indicator of the prevalence of a disease being higher (1) or lower (0) than the median prevalence across states, and the outcome is the min-max normalized value of a health score $H_{i}$. To match pairs of states, we used \emph{Propensity Score Matching} (PSM)~\cite{rosenbaum1983central}. PSM matches subjects based on a \emph{propensity score}, namely the probability of a subject being assigned to the treatment, given a set of its covariates. We obtained propensity scores by regressing the confounders to the treatment using logistic regression, and we then paired states by nearest-neighbor matching on those scores. We estimated the ATE's 95\% confidence intervals using \emph{bootstrap}: a method that assesses the variability of a measure by recalculating it on multiple re-samples of the data~\cite{bootstrappsm}. Specifically, we repeated 100 times a sampling with replacement of the set $M$ of matching pairs and calculated the confidence interval of the set of ATE values obtained for these samples.

By reviewing relevant literature, we compiled a list of possible confounders (Table~\ref{tab:conf}). For example, we included scholarization statistics based on studies that suggested a relationship between education levels and health~\cite{edu,edu2}. Overall, using open-data sources, we gathered 26 demographic, economic, social, cultural, and psychological variables defined at the state-level (Table~\ref{tab:conf}). When the number of confounders is relatively large compared to the sample size, it is preferable to reduce the set of variables to a more parsimonious set. We did so using two popular statistical approaches: the \emph{change-in-estimate}~\cite{confounder} and the \emph{High-Dimensional Propensity Score Adjustment} (HDPSA)~\cite{hdps}. The change-in-estimates selects a confounder if its inclusion changes the ATE obtained using all other covariates by a minimum threshold of $10\%$. HDPSA is a method to select the confounders whose distribution is most imbalanced between the treated and control subjects. We select our final set of confounders by intersecting the sets given in output by change-in-estimates and by HDPSA.

\begin{table}
\tiny
\begin{center}
\begin{tabular}[t]{p{35mm}p{35mm}p{5mm}}

\multicolumn{1}{c}{\textbf{\textit{Variable}}} & \multicolumn{1}{c}{\textbf{\textit{Source}}} & \multicolumn{1}{c}{\textbf{\textit{Freq}}} \\
\rowcolor{Gray} \multicolumn{3}{c}{\textbf{Demographics~\cite{demo1,demo2}}} \\ 
\% single parent households &  \multirow{3}{*}{CDC Social Vulnerability Index (SVI)} & 4\\
\% minorities estimate &  & 3\\ 
\% civilians with disability & & 1 \\
\hline
population density & \multirow{3}{*}{American Community Survey} & 6 \\
population & & 10\\
age distribution in age brackets & & 8 \\

\rowcolor{Gray}\multicolumn{3}{c}{\textbf{Economy~\cite{eco,eco2}}} \\
\% unemployed	& \multirow{3}{*}{American Community Survey} & 7\\
per capita income	&  & 5\\
official poverty measure &  & 2 \\
\hline
housing with 10+ units & \multirow{7}{*}{CDC Social Vulnerability Index (SVI)} & 1 \\
\%mobile homes &  & 2\\
housing with more ppl than rooms & & 4  \\
\% households with no vehicle &  & 2\\
\% living in group quarters	&  & 1\\
\% people below poverty estimate &   & 2\\
\% percentage uninsured &  & 6\\ \hline
gini-coefficient	& Wikipedia & 4 \\

\rowcolor{Gray}\multicolumn{3}{c}{\textbf{Education~\cite{edu,edu2}}} \\
\% people with higher education degree & American Community Survey & 4\\
\hline
\% people speaking English less than well	& Social Vulnerability Index (SVI) & 1\\

\rowcolor{Gray}\multicolumn{3}{c}{\textbf{Crime~\cite{crime}}} \\
homicide rate & \multirow{2}{*}{Wikipedia} & 3 \\
firearm death rate &  & 5\\
\hline
age-adjusted suicide rate & Center of Disease Control & 2\\

\rowcolor{Gray}\multicolumn{3}{c}{\textbf{Culture~\cite{culture}}} \\
cultural tightness & \cite{harrington2014tightness} & 5 \\
\hline
willingness of donating to charities & \multirow{2}{*}{Forbes, 2017} & 0 \\
\%people volunteering & & 3\\

\rowcolor{Gray}\multicolumn{3}{c}{\textbf{Personality~\cite{personality}}} \\
distribution over OCEAN traits & \cite{jason} & 7\\
\hline
\end{tabular}
\caption{\rev{Selection of candidate confounders (at the level of US states) based on prior literature and existing surveys. \emph{Freq} refers to the frequency with which the given candidates are selected as confounders in the causal analysis shown in Table 3.}}
\label{tab:conf}
\end{center}
\end{table}

\subsection{Causal Effect Results} \label{sec:causality:results}
We hypothesize that, after discounting the effect of relevant confounders, the increase of a health condition's prevalence causes more mentions of that condition on Reddit and, as such, a lower corresponding health score. As expected, all ATE values were negative. We reported the ATE values in Table~\ref{tab:correlations} (rightmost column), and marked the ones whose confidence intervals are entirely below zero (i.e.,~we are over 95\% confident that those ATE are negative). Among the strongest causal associations, we found that a state exhibiting levels of `mentally unhealthy days' higher than the median, after controlling for confounders, produces a 10\% decrease in the \emph{mental} health score ($H_{{mental}}$). Other noticeable effects were found for heroin use on the \emph{infection} health score (-9\%), early latent syphilis on the \emph{STDs} health score (-12\%), and for poor self-rated health on the general health scores (-12\% and -19\%). 
We also present in Table \ref{tab:conf} the frequency of candidate confounders that were selected in the causal analysis. Not surprisingly, we can observe that demographics based candidates were most likely to be selected as confounders.
In summary, in a way similar to the correlation analysis, the causal analysis corroborates both \textbf{H1} and \textbf{H2}.

\section{Discussion and conclusion}\label{sec:conclusions}

By using the lens of network science to study the co-occurrences of mentions of medical conditions in social media, we derived the first comprehensive health taxonomy from online health discussions. Its categories happen to align well with the official disease categorization. The two health taxonomies independently extracted from Reddit and Twitter are similar yet exhibit differences that reflect the different uses of the two platforms. Twitter's taxonomy focused on frequently occurring conditions, while Reddit's turned out to be more comprehensive, in that, it included less frequently occurring conditions as well. That is mainly because health discussions on Reddit are organized in specialized communities. \crc{Furthermore, our health scores computed from Reddit correlated with (and were causally linked to) the prevalence of their corresponding diseases at the level of states in the U.S.}

\crc{Our methodology is affected by several limitations that future work can address. First, even if our NLP extractor of medical conditions from text surpasses the performance of state-of-the-art solutions, its output is not a perfectly exhaustive and concise representation of all the conditions mentioned in the corpora we analyzed. In particular, our method would benefit from a better procedure of entity normalization, which would allow for grouping mentions of medical conditions with nearly-equivalent semantics. Second, our taxonomy is still a coarse representation of the complex space of all existing health conditions. This is particularly evident in some categories such as the one describing mental health---a very broad category containing thousands of terms split only among three sub-categories. Refining our taxonomy into more specific yet coherent categories below level-2 would allow for a more detailed representation of different classes of diseases and a comparison with the ICD taxonomy at the level of its sub-categories. Third, our validation is restricted by the limited number of datapoints, all representing very broad geographical areas (states in the U.S.); collecting official statistics on disease prevalence at a finer spatial granularity would alleviate this limitation. Fourth, and related to the previous point, the correlations between our health indices and official prevalence are not high. This gap can be explained by not all patients discussing their conditions online, certain states' populations being more tech-savvy, and some conditions being more likely to be discussed online than the others. The identified link in our taxonomy between measles and vaccines reveals another potential issue, which is that online discussions might correspond to perceived versus really experienced conditions. This issue, however, opens an interesting avenue for future work about potential misconceptions exhibited in online discussions, especially given the widespread scepticism over Covid-19 vaccines currently being administered. Future work could, for example, employ platform-specific signals, such as downvotes, likes, and replies to discover misconceptions among the taxonomy links. Last, like most research based on social media data, our study is affected by socio-demographic biases. Despite Reddit and Twitter's penetration is higher in the U.S. than anywhere else in the world, their user base is not representative of the population of residents; for example, it is skewed towards an audience that is younger, more affluent, and more educated than average~\cite{twitter_penetration}. This is another potential explanation for the gap between official prevalence and our indices discussed in the previous point. }

\crc{Despite these limitations, our health categorization matched the ICD-11 categorization surprisingly well at its coarsest level and the health indices we derived from it correlate with official disease prevalence statistics. However, investigating the gap between the prevalence of disease and the volume of its mentions in social media is necessary to shed light upon the relationship between which medical conditions occupy people's minds as opposed to which trouble their bodies. Shedding further light on this aspect is key to designing an integration of our health indices with official health surveys and other health surveillance systems.}

\bibliographystyle{aaai}


\begin{thebibliography}{}

\bibitem[\protect\citeauthoryear{Akbik, Blythe, and
  Vollgraf}{2018}]{akbik2018contextual}
Akbik, A.; Blythe, D.; and Vollgraf, R.
\newblock 2018.
\newblock Contextual string embeddings for sequence labeling.
\newblock In {\em Proceedings of the International Conference on Computational
  Linguistics (COLING)},  1638--1649.

\bibitem[\protect\citeauthoryear{Austin and Small}{2014}]{bootstrappsm}
Austin, P.~C., and Small, D.~S.
\newblock 2014.
\newblock {The use of bootstrapping when using propensity-score matching
  without replacement: A simulation study}.
\newblock {\em Statistics in Medicine} 33(24):4306--4319.

\bibitem[\protect\citeauthoryear{Bagroy, Kumaraguru, and
  De~Choudhury}{2017}]{bagroy2017social}
Bagroy, S.; Kumaraguru, P.; and De~Choudhury, M.
\newblock 2017.
\newblock A social media based index of mental well-being in college campuses.
\newblock In {\em Proceedings of the Conference on Human factors in Computing
  Systems (CHI)},  1634--1646.

\bibitem[\protect\citeauthoryear{Balsamo, Bajardi, and
  Panisson}{2019}]{Balsamo2019}
Balsamo, D.; Bajardi, P.; and Panisson, A.
\newblock 2019.
\newblock Firsthand opiates abuse on social media: Monitoring geospatial
  patterns of interest through a digital cohort.
\newblock In {\em Proceedings of the ACM World Wide Web Conference (WWW)},
  2572--2579.

\bibitem[\protect\citeauthoryear{Baumgartner \bgroup et al\mbox.\egroup
  }{2020}]{baumgartner2020pushshift}
Baumgartner, J.; Zannettou, S.; Keegan, B.; Squire, M.; and Blackburn, J.
\newblock 2020.
\newblock {The Pushshift Reddit Dataset}.
\newblock {\em arXiv preprint arXiv:2001.08435}.

\bibitem[\protect\citeauthoryear{Bogdashina}{2016}]{bogdashina2016sensory}
Bogdashina, O.
\newblock 2016.
\newblock {\em {Sensory perceptual issues in autism and Asperger syndrome:
  Different sensory experiences-different perceptual worlds}}.
\newblock Jessica Kingsley Publishers.

\bibitem[\protect\citeauthoryear{Boldi and Vigna}{2014}]{boldi2014axioms}
Boldi, P., and Vigna, S.
\newblock 2014.
\newblock Axioms for centrality.
\newblock {\em Internet Mathematics} 10(3-4):222--262.

\bibitem[\protect\citeauthoryear{Booth-Kewley and
  Vickers~Jr}{1994}]{personality}
Booth-Kewley, S., and Vickers~Jr, R.~R.
\newblock 1994.
\newblock Associations between major domains of personality and health
  behavior.
\newblock {\em Journal of Personality} 62(3):281--298.

\bibitem[\protect\citeauthoryear{Butler}{2013}]{butler13}
Butler, D.
\newblock 2013.
\newblock {When Google Got Flu Wrong}.
\newblock {\em Nature} 494:155--6.

\bibitem[\protect\citeauthoryear{Chan \bgroup et al\mbox.\egroup
  }{2018}]{chan2018online}
Chan, M.-p.~S.; Lohmann, S.; Morales, A.; Zhai, C.; Ungar, L.; Holtgrave,
  D.~R.; and Albarrac{\'\i}n, D.
\newblock 2018.
\newblock {An online risk index for the cross-sectional prediction of new HIV
  chlamydia, and gonorrhea diagnoses across US counties and across years}.
\newblock {\em AIDS and Behavior} 22(7):2322--2333.

\bibitem[\protect\citeauthoryear{Choudhury and De}{2014}]{de2014mental}
Choudhury, M.~D., and De, S.
\newblock 2014.
\newblock Mental health discourse on reddit: Self-disclosure, social support,
  and anonymity.
\newblock In {\em Proceedings of the International AAAI Conference on Weblogs
  and Social Media (ICWSM)}.

\bibitem[\protect\citeauthoryear{Cochrane, OHara, and Leslie}{1980}]{edu}
Cochrane, S.~H.; OHara, D.~J.; and Leslie, J.
\newblock 1980.
\newblock The effects of education on health.
\newblock Technical Report SWP405, Washington DC World Bank.

\bibitem[\protect\citeauthoryear{Coscia and Neffke}{2017}]{coscia2017network}
Coscia, M., and Neffke, F.~M.
\newblock 2017.
\newblock Network backboning with noisy data.
\newblock In {\em 2017 IEEE 33rd International Conference on Data Engineering
  (ICDE)},  425--436.
\newblock IEEE.

\bibitem[\protect\citeauthoryear{Culotta}{2014}]{Culotta204twitter}
Culotta, A.
\newblock 2014.
\newblock Estimating county health statistics with twitter.
\newblock In {\em Proceedings of the ACM Conference on Human Factors in
  Computing Systems (CHI)},  1335--1344.

\bibitem[\protect\citeauthoryear{Cutler and Lleras-Muney}{2006}]{edu2}
Cutler, D.~M., and Lleras-Muney, A.
\newblock 2006.
\newblock {Education and health: Evaluating theories and evidence}.
\newblock Technical report, National Bureau of Economic Research.

\bibitem[\protect\citeauthoryear{De~Choudhury \bgroup et al\mbox.\egroup
  }{2013}]{de2013predicting}
De~Choudhury, M.; Gamon, M.; Counts, S.; and Horvitz, E.
\newblock 2013.
\newblock Predicting depression via social media.
\newblock In {\em Proceedings of the International AAAI conference on Weblogs
  and Social Media (ICWSM)}.

\bibitem[\protect\citeauthoryear{Deaton}{2008}]{eco2}
Deaton, A.
\newblock 2008.
\newblock Income, health, and well-being around the world: Evidence from the
  gallup world poll.
\newblock {\em Journal of Economic Perspectives} 22(2):53--72.

\bibitem[\protect\citeauthoryear{Devlin \bgroup et al\mbox.\egroup
  }{2019}]{devlin2019bert}
Devlin, J.; Chang, M.-W.; Lee, K.; and Toutanova, K.
\newblock 2019.
\newblock Bert: Pre-training of deep bidirectional transformers for language
  understanding.
\newblock In {\em Proceedings of the Conference of the North American Chapter
  of the Association for Computational Linguistics: Human Language
  Technologies},  4171--4186.

\bibitem[\protect\citeauthoryear{Fortunato}{2010}]{fortunato2010community}
Fortunato, S.
\newblock 2010.
\newblock Community detection in graphs.
\newblock {\em Physics Reports} 486(3-5):75--174.

\bibitem[\protect\citeauthoryear{Gillam, Siriwardena, and Steel}{2012}]{Gillam}
Gillam, S.~J.; Siriwardena, A.~N.; and Steel, N.
\newblock 2012.
\newblock {Pay-for-performance in the United Kingdom: impact of the quality and
  outcomes framework: a systematic review}.
\newblock {\em {Annals of Family Medicine}} 10:461--68.

\bibitem[\protect\citeauthoryear{Gkotsis \bgroup et al\mbox.\egroup
  }{2017}]{gkotsis2017characterisation}
Gkotsis, G.; Oellrich, A.; Velupillai, S.; Liakata, M.; Hubbard, T.~J.; Dobson,
  R.~J.; and Dutta, R.
\newblock 2017.
\newblock Characterisation of mental health conditions in social media using
  informed deep learning.
\newblock {\em Scientific Reports} 7:45141.

\bibitem[\protect\citeauthoryear{Goel \bgroup et al\mbox.\egroup
  }{2010}]{Goel17486}
Goel, S.; Hofman, J.~M.; Lahaie, S.; Pennock, D.~M.; and Watts, D.~J.
\newblock 2010.
\newblock {Predicting consumer behavior with Web search}.
\newblock {\em {Proceedings of the National Academy of Sciences (PNAS)}}
  107(41):17486--17490.

\bibitem[\protect\citeauthoryear{Goh \bgroup et al\mbox.\egroup
  }{2007}]{goh2007human}
Goh, K.-I.; Cusick, M.~E.; Valle, D.; Childs, B.; Vidal, M.; and Barab{\'a}si,
  A.-L.
\newblock 2007.
\newblock The human disease network.
\newblock {\em Proceedings of the National Academy of Sciences (PNAS)}
  104(21):8685--8690.

\bibitem[\protect\citeauthoryear{Greenland}{2008}]{confounder}
Greenland, S.
\newblock 2008.
\newblock Invited commentary: variable selection versus shrinkage in the
  control of multiple confounders.
\newblock {\em American Journal of Epidemiology} 167(5):523--529.

\bibitem[\protect\citeauthoryear{Guntuku \bgroup et al\mbox.\egroup
  }{2019}]{guntuku2019understanding}
Guntuku, S.~C.; Buffone, A.; Jaidka, K.; Eichstaedt, J.~C.; and Ungar, L.~H.
\newblock 2019.
\newblock Understanding and measuring psychological stress using social media.
\newblock In {\em Proceedings of the International AAAI Conference on Web and
  Social Media (ICWSM)},  214--225.

\bibitem[\protect\citeauthoryear{Harrington and
  Gelfand}{2014}]{harrington2014tightness}
Harrington, J.~R., and Gelfand, M.~J.
\newblock 2014.
\newblock Tightness--looseness across the 50 united states.
\newblock {\em Proceedings of the National Academy of Sciences (PNAS)}
  111(22):7990--7995.

\bibitem[\protect\citeauthoryear{Huang, Xu, and
  Yu}{2015}]{huang2015bidirectional}
Huang, Z.; Xu, W.; and Yu, K.
\newblock 2015.
\newblock Bidirectional lstm-crf models for sequence tagging.
\newblock {\em arXiv preprint arXiv:1508.01991}.

\bibitem[\protect\citeauthoryear{Jiang \bgroup et al\mbox.\egroup
  }{2019}]{jiang2019improved}
Jiang, Y.; Hu, C.; Xiao, T.; Zhang, C.; and Zhu, J.
\newblock 2019.
\newblock Improved differentiable architecture search for language modeling and
  named entity recognition.
\newblock In {\em Proceedings of the Conference on Empirical Methods in Natural
  Language Processing and the International Joint Conference on Natural
  Language Processing (EMNLP-IJCNLP)},  3576--3581.

\bibitem[\protect\citeauthoryear{Jimeno-Yepes \bgroup et al\mbox.\egroup
  }{2015}]{jimeno2015identifying}
Jimeno-Yepes, A.; MacKinlay, A.; Han, B.; and Chen, Q.
\newblock 2015.
\newblock Identifying diseases, drugs, and symptoms in twitter.
\newblock {\em Studies in Health Technology and Informatics} 216:643.

\bibitem[\protect\citeauthoryear{Karimi \bgroup et al\mbox.\egroup
  }{2015}]{karimi2015cadec}
Karimi, S.; Metke-Jimenez, A.; Kemp, M.; and Wang, C.
\newblock 2015.
\newblock Cadec: A corpus of adverse drug event annotations.
\newblock {\em Journal of Biomedical Informatics} 55:73--81.

\bibitem[\protect\citeauthoryear{Kass-Hout and
  Alhinnawi}{2013}]{kass2013social}
Kass-Hout, T., and Alhinnawi, H.
\newblock 2013.
\newblock Social media in public health.
\newblock {\em British Medical Bulletin} 108:5.

\bibitem[\protect\citeauthoryear{Kramer}{2010}]{kramer2010unobtrusive}
Kramer, A.~D.
\newblock 2010.
\newblock An unobtrusive behavioral model of gross national happiness.
\newblock In {\em Proceedings of the SIGCHI Conference on Human Factors in
  Computing Systems (CHI)},  287--290.
\newblock ACM.

\bibitem[\protect\citeauthoryear{Lancichinetti and
  Fortunato}{2009}]{lancichinetti2009community}
Lancichinetti, A., and Fortunato, S.
\newblock 2009.
\newblock Community detection algorithms: a comparative analysis.
\newblock {\em Physical review E} 80(5):056117.

\bibitem[\protect\citeauthoryear{Lawson \bgroup et al\mbox.\egroup
  }{2010}]{lawson2010annotating}
Lawson, N.; Eustice, K.; Perkowitz, M.; and Yetisgen-Yildiz, M.
\newblock 2010.
\newblock {Annotating large email datasets for named entity recognition with
  Mechanical Turk}.
\newblock In {\em Proceedings of the ACM NAACL HLT Workshop on Creating Speech
  and Language Data with Amazon's Mechanical Turk},  71--79.

\bibitem[\protect\citeauthoryear{Lazer \bgroup et al\mbox.\egroup
  }{2014}]{Lazer1203}
Lazer, D.; Kennedy, R.; King, G.; and Vespignani, A.
\newblock 2014.
\newblock {The Parable of Google Flu: Traps in Big Data Analysis}.
\newblock {\em {Science}} 343(6176):1203--1205.

\bibitem[\protect\citeauthoryear{Liu \bgroup et al\mbox.\egroup
  }{2019}]{liu2019roberta}
Liu, Y.; Ott, M.; Goyal, N.; Du, J.; Joshi, M.; Chen, D.; Levy, O.; Lewis, M.;
  Zettlemoyer, L.; and Stoyanov, V.
\newblock 2019.
\newblock {RoBERTa: A robustly optimized BERT pre-training approach}.
\newblock {\em arXiv preprint arXiv:1907.11692}.

\bibitem[\protect\citeauthoryear{McAfee and
  Brynjolfsson}{2012}]{mcafee12bigdata}
McAfee, A., and Brynjolfsson, E.
\newblock 2012.
\newblock {Big Data}: The management revolution.
\newblock {\em {Harvard Business Review}} 90(10):60--68.

\bibitem[\protect\citeauthoryear{Medvedev, Lambiotte, and
  Delvenne}{2017}]{medvedev2017anatomy}
Medvedev, A.~N.; Lambiotte, R.; and Delvenne, J.-C.
\newblock 2017.
\newblock {The anatomy of Reddit: An overview of academic research}.
\newblock In {\em Dynamics on and of Complex Networks},  183--204.
\newblock Springer.

\bibitem[\protect\citeauthoryear{Meer, Miller, and Rosen}{2003}]{eco}
Meer, J.; Miller, D.~L.; and Rosen, H.~S.
\newblock 2003.
\newblock Exploring the health--wealth nexus.
\newblock {\em Journal of Health Economics} 22(5):713--730.

\bibitem[\protect\citeauthoryear{Napier \bgroup et al\mbox.\egroup
  }{2014}]{culture}
Napier, A.~D.; Ancarno, C.; Butler, B.; Calabrese, J.; Chater, A.; Chatterjee,
  H.; Guesnet, F.; Horne, R.; Jacyna, S.; Jadhav, S.; et~al.
\newblock 2014.
\newblock Culture and health.
\newblock {\em The Lancet} 384(9954):1607--1639.

\bibitem[\protect\citeauthoryear{Neale and Kendler}{1995}]{neale1995models}
Neale, M.~C., and Kendler, K.~S.
\newblock 1995.
\newblock Models of comorbidity for multifactorial disorders.
\newblock {\em American Journal of Human Genetics} 57(4):935.

\bibitem[\protect\citeauthoryear{Page \bgroup et al\mbox.\egroup
  }{1999}]{page1999pagerank}
Page, L.; Brin, S.; Motwani, R.; and Winograd, T.
\newblock 1999.
\newblock {The PageRank citation ranking: Bringing order to the web}.
\newblock Technical report, Stanford InfoLab.

\bibitem[\protect\citeauthoryear{Pennebaker, Francis, and
  Booth}{2001}]{pennebaker2001linguistic}
Pennebaker, J.~W.; Francis, M.~E.; and Booth, R.~J.
\newblock 2001.
\newblock Linguistic inquiry and word count: Liwc 2001.
\newblock {\em Mahway: Lawrence Erlbaum Associates} 71.

\bibitem[\protect\citeauthoryear{Perrin and
  Anderson}{2018}]{twitter_penetration}
Perrin, A., and Anderson, M.
\newblock 2018.
\newblock Share of {U.S.} adults using social media, including facebook, is
  mostly unchanged since 2018.

\bibitem[\protect\citeauthoryear{Rentfrow, Gosling, and Potter}{2008}]{jason}
Rentfrow, P.~J.; Gosling, S.~D.; and Potter, J.
\newblock 2008.
\newblock A theory of the emergence, persistence, and expression of geographic
  variation in psychological characteristics.
\newblock {\em Perspectives on Psychological Science} 3(5):339--369.

\bibitem[\protect\citeauthoryear{Rosenbaum and
  Rubin}{1983}]{rosenbaum1983central}
Rosenbaum, P.~R., and Rubin, D.~B.
\newblock 1983.
\newblock The central role of the propensity score in observational studies for
  causal effects.
\newblock {\em Biometrika} 70(1):41--55.

\bibitem[\protect\citeauthoryear{Rosvall and Bergstrom}{2008}]{rosvall2008maps}
Rosvall, M., and Bergstrom, C.~T.
\newblock 2008.
\newblock Maps of random walks on complex networks reveal community structure.
\newblock {\em Proceedings of the National Academy of Sciences (PNAS)}
  105(4):1118--1123.

\bibitem[\protect\citeauthoryear{Sarasohn-Kahn}{2008}]{sarasohn2008wisdom}
Sarasohn-Kahn, J.
\newblock 2008.
\newblock {The wisdom of patients: Health care meets online social media}.
\newblock Technical report, California HealthCare Foundation.

\bibitem[\protect\citeauthoryear{Schneeweiss \bgroup et al\mbox.\egroup
  }{2009}]{hdps}
Schneeweiss, S.; Rassen, J.~A.; Glynn, R.~J.; Avorn, J.; Mogun, H.; and
  Brookhart, M.~A.
\newblock 2009.
\newblock High-dimensional propensity score adjustment in studies of treatment
  effects using health care claims data.
\newblock {\em Epidemiology (Cambridge, Mass.)} 20(4):512.

\bibitem[\protect\citeauthoryear{Stafford, Chandola, and Marmot}{2007}]{crime}
Stafford, M.; Chandola, T.; and Marmot, M.
\newblock 2007.
\newblock Association between fear of crime and mental health and physical
  functioning.
\newblock {\em American Journal of Public Health} 97(11):2076--2081.

\bibitem[\protect\citeauthoryear{Stephens-Davidowitz}{2018}]{seth18everybody}
Stephens-Davidowitz, S.
\newblock 2018.
\newblock {\em {Everybody Lies: Big Data, New Data, and What the Internet Can
  Tell Us About Who We Really Are}}.
\newblock William Morrow \& Co.

\bibitem[\protect\citeauthoryear{Stordal \bgroup et al\mbox.\egroup
  }{2001}]{demo1}
Stordal, E.; Bjartveit~Kr{\"u}ger, M.; Dahl, N.~H.; Kr{\"u}ger, {\O}.;
  Mykletun, A.; and Dahl, A.
\newblock 2001.
\newblock {Depression in relation to age and gender in the general population:
  The Nord-Tr{\o}ndelag Health Study (HUNT)}.
\newblock {\em Acta Psychiatrica Scandinavica} 104(3):210--216.

\bibitem[\protect\citeauthoryear{Taylor \bgroup et al\mbox.\egroup
  }{1999}]{taylor1999autism}
Taylor, B.; Miller, E.; Farrington, C.; Petropoulos, M.-C.; Favot-Mayaud, I.;
  Li, J.; and Waight, P.~A.
\newblock 1999.
\newblock Autism and measles, mumps, and rubella vaccine: no epidemiological
  evidence for a causal association.
\newblock {\em The Lancet} 353(9169):2026--2029.

\bibitem[\protect\citeauthoryear{Toussaint \bgroup et al\mbox.\egroup
  }{2001}]{demo2}
Toussaint, L.~L.; Williams, D.~R.; Musick, M.~A.; and Everson, S.~A.
\newblock 2001.
\newblock Forgiveness and health: Age differences in a us probability sample.
\newblock {\em Journal of Adult Development} 8(4):249--257.

\bibitem[\protect\citeauthoryear{Tutubalina and
  Nikolenko}{2017}]{tutubalina2017combination}
Tutubalina, E., and Nikolenko, S.
\newblock 2017.
\newblock Combination of deep recurrent neural networks and conditional random
  fields for extracting adverse drug reactions from user reviews.
\newblock {\em Journal of Healthcare Engineering} 2017.

\bibitem[\protect\citeauthoryear{Velasco \bgroup et al\mbox.\egroup
  }{2014}]{velasco2014social}
Velasco, E.; Agheneza, T.; Denecke, K.; Kirchner, G.; and Eckmanns, T.
\newblock 2014.
\newblock Social media and internet-based data in global systems for public
  health surveillance: a systematic review.
\newblock {\em The Milbank Quarterly} 92(1):7--33.

\bibitem[\protect\citeauthoryear{Yepes and MacKinlay}{2016}]{yepes2016ner}
Yepes, A.~J., and MacKinlay, A.
\newblock 2016.
\newblock {NER} for medical entities in {T}witter using sequence to sequence
  neural networks.
\newblock In {\em Proceedings of the Australasian Language Technology
  Association Workshop},  138--142.

\bibitem[\protect\citeauthoryear{Zhou \bgroup et al\mbox.\egroup
  }{2014}]{zhou2014human}
Zhou, X.; Menche, J.; Barab{\'a}si, A.-L.; and Sharma, A.
\newblock 2014.
\newblock Human symptoms--disease network.
\newblock {\em Nature Communications} 5:4212.

\end{thebibliography}

\end{document}